\documentclass{article}
\usepackage{multirow} % 用于合并多行单元格
\usepackage{booktabs} % 提供更美观的横线命令
\usepackage{makecell}  % 用于单元格内换行（可选，此处用于优化列标题）
\usepackage{graphicx}  
\usepackage{caption}    
\usepackage{subcaption} 
\graphicspath{{figures/}}  
\usepackage{dashrule}      
\usepackage{geometry}     \usepackage{amssymb}  % 提供 \mathbb 等数学符号命令
\usepackage{spconf,amsmath,graphicx,hyperref}

\makeatletter
\renewcommand{\@makefnmark}{\hbox{\@textsuperscript{\fnsymbol{footnote}}}}
\makeatother
\title{CPCLDETECTOR: Knowledge Enhancement and Alignment Selection for Chinese Patronizing and Condescending Language Detection}
\name{Jiaxun Yang$^{1}$, Yifei Han$^2$, Long Zhang$^2$, Yujie Liu$^{3}$, Bin Li$^{2\textasteriskcentered}$, Bo Gao$^{3\textasteriskcentered}$, Yangfan He$^{4}$, Kejia Zhang$^{5}$}

\vspace{-3em}

\address{$^1$City University of Hong Kong,  $^2$SIAT, CAS \\
$^3$School of Information Engineering, Beijing Institute of Graphic Communication,
\\ $^4$University of Minnesota - Twin Cities, $^5$Heilongjiang University}
\begin{document}
%\ninept
%
\maketitle

\begin{abstract}
Chinese Patronizing and Condescending Language (CPCL) is an implicitly discriminatory toxic speech targeting vulnerable groups on Chinese video platforms. The existing dataset lacks user comments, which are a direct reflection of video content. This undermines the model's understanding of video content and results in the failure to detect some CPLC videos. To make up for this loss, this research reconstructs a new dataset PCLMMPLUS that includes 103k comment entries and expands the dataset size. We also propose the CPCLDetector model with alignment selection and knowledge-enhanced comment content modules. Extensive experiments show the proposed CPCLDetector outperforms the SOTA on PCLMM and achieves higher performance on PCLMMPLUS . CPLC videos are detected more accurately, supporting content governance and protecting vulnerable groups. Code and dataset are available at https://github.com/jiaxunyang256/PCLD.
\end{abstract}
\begin{keywords}
Chinese Patronizing and Condescending Language (CPCL), Multi-Modal Detection, Alignment, Knowledge Enhancement.
\end{keywords}
\vspace{-0.3cm}
\section{Introduction}
\vspace{-0.1cm}
\label{sec:intro}
\footnotetext{$^{*}$Corresponding author.}
With the continuous development of video platforms in
China, the content communities represented by Douyin and Bilibili have become important platforms for the dissemination of public content~\cite{diao2023av,li2023overview,li2024towards,diao2025temporal}. However, accurate governance of such information has become a core demand in both industry and academia. Currently, research on toxic speech focuses mainly on explicit harmful content such as hate speech. Although relevant technical means can control the spread of such content to a certain extent, there is still insufficient attention paid to another type of implicitly discriminatory speech—Patronizing and Condescending Language (PCL)~\cite{perez2020don}.

As a type of discriminatory toxic speech targeting vulnerable groups, PCL is characterized by implicit expression. It does not contain obvious offensive words, but conveys a sense of "superiority" through a hypocritical and condescending attitude, causing harm to specific groups. The trend of user comments can reflect the orientation of a video ~\cite{obadimu2019identifying}~\cite{miyazaki2024impact}~\cite{Bertaglia_2024}. Wang et al.~\cite{wang2025towards}
constructed the first multi-modal dataset for Chinese video
PCL detection (PCLMM) and the MultiPCL detection model,
providing fundamental support for this field. 

This study focuses on CPCL detection in video and performs two core tasks:
Constructing the Chinese PCL Multi-modal Dataset
PCLMMPLUS: Based on the PCLMM dataset, the sample
size is expanded to 831 valid video samples (with a total duration of more than 26 hours), and user comment data corresponding  
to the videos are incorporated to provide more comprehensive
data support for CPCL detection. Designing the Multi-modal
Adaptive CPCL Detection Model CPCLDetector: Based on
the PCLMMPLUS dataset, The model includes alignment selection and knowledge-enhanced comment content modules.
In the PCLMM data set, the CPCLDetector demonstrates superior performance compared to the existing MultiPCL model. It also shows excellent performance on the PCLMMPLUS dataset, proving that supplementary modules such as the knowledge-enhanced comment content module can further improve the detection accuracy. 
\par
The \textbf{main contributions} of this paper can be summarized as follows:
(1) Builds the PCLMMPLUS dataset. It expands the dataset size and includes 103k user comments, filling the comment modality blank. (2) Proposes the CPCLDetector model. It integrates two core modules—alignment selection and knowledge-enhanced comment content—to leverage multi-modal information for better CPCL detection. (3) Extensive experiments show that our method outperforms SOTA methods on the PCLMM dataset and achieves higher results on the PCLMMPLUS dataset.  
\vspace{-0.4cm}
\section{DATASET}
\label{sec:format}
\vspace{-0.1cm}
\subsection{Research Background of Datasets}

In the research on CPCL detection, the lack of multi-modal data has long restricted technological development. Existing achievements mostly focus on the English field~\cite{wang2019talkdown}~\cite{perez2020don} or single Chinese text modality~\cite{wang2023ccpc} , making it difficult to adapt to the characteristics of multi-modal CPCL intent detection in Chinese video scenarios. To solve this problem, the PCLMM dataset was constructed. On this basis, this study further expands it to build the PCLMMPLUS dataset. By expanding the sample size and adding new information dimensions, the supporting capacity of the dataset for CPCL detection research is further enhanced.
\vspace{-0.3cm}
\begin{table}[htbp]  % 单栏表格环境
    \centering
    \caption{Comparison between the prior PCLMM~~\cite{perez2020don} and the proposed CPCLMMPLUS.}  % 表格标题（必加，方便引用）
    \large  % 全局放大字体（可选 \Large 进一步放大，根据需求调整）
    % 缩减列宽（从1.5cm→1.1cm），减少横向占用，避免字体被过度压缩
    \resizebox{\linewidth}{!}{  % 仅轻微缩放以适配单栏，优先保证字体大小
        \begin{tabular}{c *{6}{>{\centering\arraybackslash}p{1.1cm}}} 
            \toprule % 顶部粗线
            \multirow{2}{*}{Statistics} & \multicolumn{3}{c}{PCLMM~~\cite{wang2025towards}} & \multicolumn{3}{c}{CPCLMMPLUS (Ours)} \\
            \cmidrule(lr){2-4} \cmidrule(lr){5-7} % 子列分隔线（避免与文本重叠）
            & Non-PCL & PCL & Total & Non-PCL & PCL & Total\\
            \midrule % 中间线
            Videos total num & 519 & 196 & 715 & 519 & 312 & 831 \\
            Videos total len (hrs) & 15.1 & 6.5 & 21.6 & 15.1 & 11.9 & 27 \\
            Comments total num & - & - & - & 77K & 26K & 103K \\
            Comments total len (chars) & - & - & - & 2935K & 992K & 3927K \\
            \bottomrule % 底部粗线
        \end{tabular}
    }
    \label{Table 1}  % label放在caption后，确保引用有效
\end{table}
% \begin{table*}[t] % [t]指定表格位于页顶，*表示横跨两栏
%   \centering
%   \small % 适当缩小字体，避免过宽
%   \caption{Statistics of PCLMM and PCLMPLUS} % 表格标题
%   \label{tab:database_comparison} % 引用标签
%   % 列格式：第一列居中，后六列固定宽度1.2cm且居中
%   \begin{tabular}{c *{6}{>{\centering\arraybackslash}p{1.5cm}}} 
%     \toprule % 顶部粗线
%     \multirow{2}{*}{Database} & \multicolumn{3}{c}{PCMM} & \multicolumn{3}{c}{PCLMPLUS} \\
%     \cmidrule(lr){2-4} \cmidrule(lr){5-7} % 子列分隔线，(lr)避免与文本重叠
%     & Non-PCL & PCL & Total & Non-PCL & PCL & Total\\
%     \midrule % 中间线
%     Videos total num & 519 & 196 & 715 & 519 & 312 & 831 \\
%     Videos total len (hrs) & 15.1 & 6.5 & 21.6 & 15.1 & 11.9 & 27 \\
%     Comments total num & - & - & - & 76754 & 26386 & 103140 \\
%     Comments total len (chars) & - & - & - & 2935068 & 992373 & 3927441 \\
%     \bottomrule % 底部粗线
%   \end{tabular}
% \label{Table 1}
% \end{table*}
\vspace{-0.4cm}
\begin{figure*}[h]
	\centering
    \includegraphics[width=0.8\linewidth]{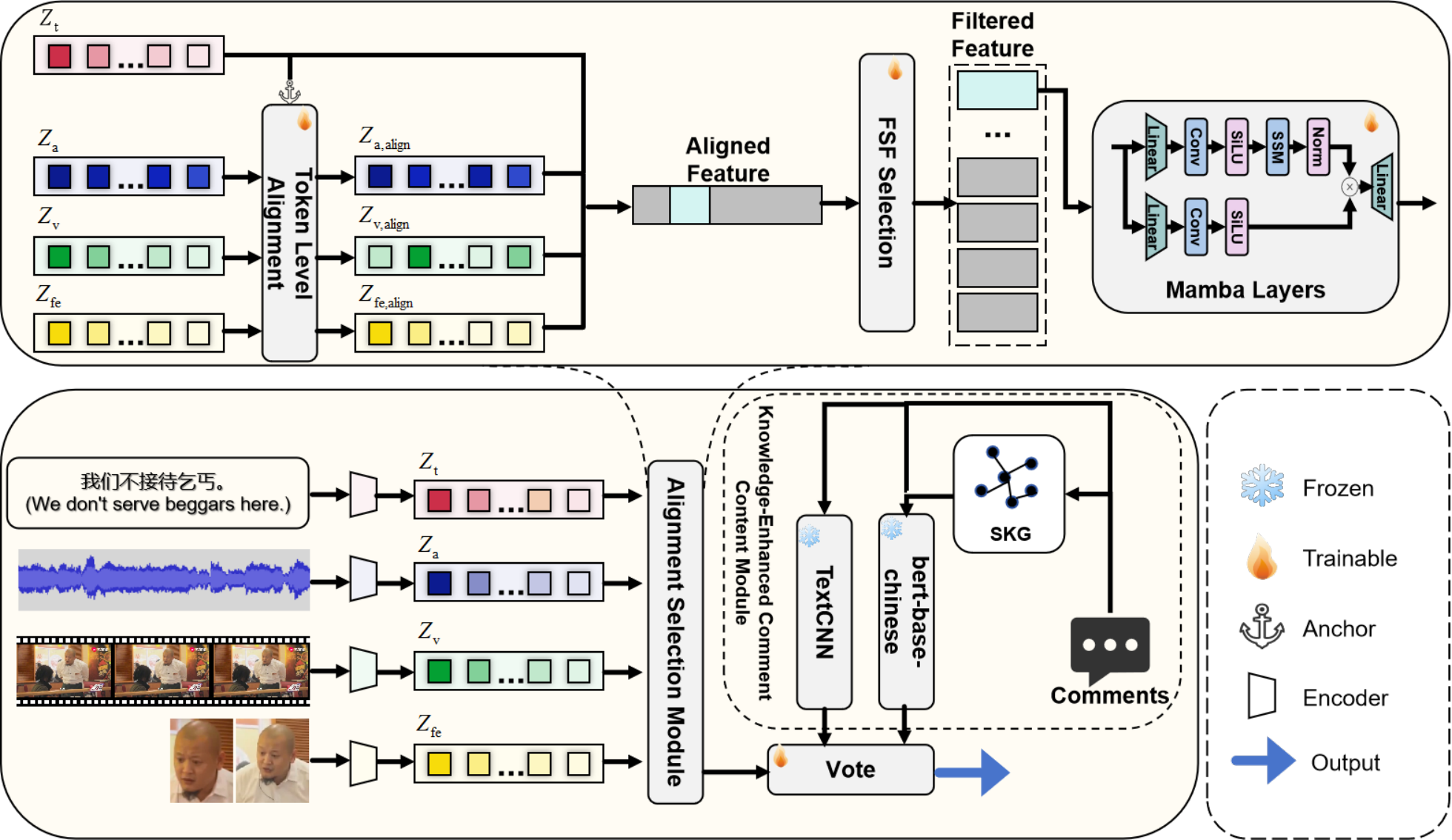}
    \vspace{-0.3cm}
	\caption{Overview of the CPCLDetector Framework, showing details of the alignment selection and knowledge-enhanced comment content modules.}
    \vspace{-0.5cm}
	\label{fig1}
\end{figure*}
\vspace{-0.5cm}
\subsection{The PCLMM Dataset}
The PCLMM dataset~\cite{wang2025towards} is the first multi-modal dataset for CPCL detection in Chinese videos targeting 6 vulnerable groups in China  from Bilibili. The dataset contains 715 valid annotated samples with 196 PCL samples and 519 non-PCL samples were obtained.
\vspace{-0.3cm}
\subsection{The PCLMMPLUS Dataset}
During the research, it was found that the proportion on CPCL samples in the PCLMM dataset is too small, resulting in models trained on this dataset having high detection accuracy for non-PCL videos but significantly lower detection performance for CPCL videos. To address the limitations of the PCLMM dataset in terms of PCL sample size and information dimensions, this study expands it based on PCLMM to construct the PCLMMPLUS dataset. This dataset inherits the core design logic of PCLMM while achieving dual upgrades in sample quantity and information richness, adding 116 new PCL video samples with a total duration of 5.4 hours. 

\textbf{Addition of Information Dimensions.} The core innovation of the PCLMMPLUS dataset lies in the addition of user comment data corresponding to each video. The research team obtained public comments of videos through the platform's compliant interface. Only first-level comments were selected here. The comments were processed through a data cleaning process, including deduplication, filtering out comments with meaningless characters, and removing comments composed solely of emojis. 
\vspace{-0.3cm}
\section{Method}
\label{sec:pagestyle}

To classify video samples V in the dataset into PCL videos (label = 1) and non-PCL videos (label = 0) based on whether they contain CPCL content, this study proposes a multi-modal model. This model focuses on aligning the multi-modal features of the video itself, processes text, audio, video, and facial features, achieves feature consistency through multi-modal difference loss, and integrates sentiment and comment information to assist detection. The overall architecture of CPCLDetector is shown in Fig. 1, and the components of the framework and training objectives are described in detail below.
\vspace{-0.3cm}
\subsection{Alignment Selection Module}
The alignment selection module aims to unify the spatio-temporal dimensions of multi-source modal features (video, facial expressions, audio, text), reduce cross-modal distribution differences, and lay a foundation for efficient multi-modal fusion.
\vspace{-0.3cm}
\subsubsection{Video Encoding}
Let the video frame sequence be\(F = \{f_1, f_2, \dots, f_n\}\). A Vision Transformer (ViT) ~\cite{dosovitskiy2020image}is used to map each frame image to a feature vector of fixed dimension. The calculation formula is as follows:
\begin{equation}
 z_{v,i} = \text{ViT}(f_i), \quad z_{v,i} \in \mathbb{R}^d
\end{equation}
where \(d = 768\), and \(z_{v,i}\) represents the feature vector of the i-th frame image.
\vspace{-0.3cm}
\subsubsection{Facial Expression Encoding}
A Multi-task Cascaded Convolutional Network (MTCNN) ~\cite{zhang2016joint}is used for face detection. If a face is detected, a facial expression recognition model based on Vision Transformer (FER-VT)~\cite{huang2021facial} is used to extract facial features. The calculation formula is as follows:
\begin{equation}
z_{fe,i} = \begin{cases} \text{FER-VT}(f_{fe,i}) & \text{if a face is detected} \\ \mathbf{0} \in \mathbb{R}^d & \text{if no face is detected} \end{cases}
\end{equation}
where \(\mathbf{0} \in \mathbb{R}^d\) represents a d-dimensional zero vector.
\vspace{-0.3cm}
\subsubsection{Audio Encoding}
The audio in the video is extracted using the multimedia toolkit FFmpeg~\cite{Ffmpeg}, and the Mel-Frequency Cepstral Coefficients (MFCC) are used to extract audio features. Finally, a feature sequence is obtained:
\begin{equation}
Z_a = \{z_{a,1}, \dots, z_{a,m}\}
\end{equation}
where \(z_{a,k} \in \mathbb{R}^d\) ( \(z_{a,k}\) represents a d-dimensional audio feature vector).
\vspace{-0.3cm}
\subsubsection{Text Encoding}
The audio is transcribed into text using the Whisper$^1$\footnotetext{$^{1}$https://github.com/openai/whisper.} speech recognition model, and then the text features are extracted using the RoBERTa-Chinese model~\cite{cui2020revisiting}. The calculation formula is as follows:
\begin{equation}
    Z_t = \text{RoBERTa-Chinese}(T), \quad Z_t = \{z_{t,1}, \dots, z_{t,q}\}
\end{equation}
where T represents the transcribed text, \(Z_t\) is the extracted text feature sequence, and \(z_{t,1},\dots,z_{t,q}\) are all d-dimensional feature vectors.
% \vspace{-0.3cm}
\subsubsection{Multi-modal Fusion}
With text as the benchmark, audio, video, and facial features are aligned at the token level using text features as anchors through a Relaxed Optimal Transport matrix. The total Maximum Mean Discrepancy (MMD) loss is calculated, which is the sum of the distribution differences between each aligned modality and the text modality~\cite{li2025alignmamba} . The calculation formula is as follows:
%\begin{equation}
\begin{align}\label{equ}
\nonumber
\mathcal{L}_{\text{MMD}} &= \text{MMD}(Z_{a,\text{align}}, Z_t) \\&+ \text{MMD}(Z_{v,\text{align}}, Z_t) + \text{MMD}(Z_{fe,\text{align}}, Z_t)
\end{align}
%\end{equation}
where \(Z_{a,\text{align}}\), \(Z_{v,\text{align}}\), and \(Z_{fe,\text{align}}\) represent the aligned audio, video, and facial feature sequences respectively, and \(\text{MMD}(\cdot)\) represents the Maximum Mean Discrepancy function. The aligned multi-modal features are concatenated, and the concatenated sequence is projected to the target dimension through a linear layer with layer normalization. Subsequently, feature selection is performed through Feature Selection Fusion (FSF) ~\cite{ma2024less} , and finally, the sequence is input into the Mamba encoder to model long-range dependencies, thus completing the multi-modal fusion process.
\vspace{-0.4cm}
\subsection{Knowledge-Enhanced Comment Content Module}
\subsubsection{Knowledge-Enhanced Sentiment Analysis Module}
This module is based on the pre-trained BERT model (bert-base-chinese) and integrates information from an external Sentiment Knowledge Graph (SKG)~\cite{zhao2021knowledge}. The SKG contains sentiment triples (attribute words, sentiment words, polarity) extracted from the Chinese PCL corpus (CCPC)~\cite{wang2023ccpc}. After segmenting the comments, word embeddings are calculated using the Sentence-Transformer model; relevant triples are matched via cosine similarity, and after embedding integration, the sentiment probability distribution is obtained through the BERT model. 
\vspace{-0.3cm}
\subsubsection{Comment Information Processing Module}
Comments are cleaned, deduplicated, filtered to remove meaningless content, segmented into words, aggregated into fixed-length sequences, mapped to index sequences via a vocabulary, and processed with TextCNN to obtain comment-based CPCL probability.
\vspace{-0.3cm}
\subsection{Loss Function}
The total loss function integrates classification loss and alignment loss. The calculation formula is as follows:
\begin{equation}
    \mathcal{L}_{\text{total}} = \mathcal{L}_{\text{focal}} + \lambda \cdot \mathcal{L}_{\text{MMD}}
\end{equation}
where:\(\mathcal{L}_{\text{focal}}\) is Focal Loss, used to address the class imbalance problem.\(\mathcal{L}_{\text{MMD}}\) is the multi-modal alignment loss, used to ensure the consistency of cross-modal features. \(\lambda\) is the weight coefficient of the loss term (\(\lambda\) = 0.3).
\vspace{-0.3cm}
\section{Experiments}
\label{sec:pagestyle}

To verify the effectiveness of CPCLDetector, the research team conducted comprehensive experiments on PCLMM and PCLMMPLUS. The experiments focused on baseline model comparison, multi-modal contribution analysis, and ablation experiments, with specific details as follows. Experiments were repeated 5 times with different random seeds to ensure statistical reliability, with mean values calculated.
% 表格部分
\vspace{-0.3cm}
\begin{table}[t!]
  \centering
  \small
  \caption{Comparison of Different Models' Performance on the PCLMM.}
  \label{tab:model_performance_new}
  \begin{tabular}{>{\centering}p{2cm}|c|c|c|c}  % 5列，格式正确
    \hline
    \textbf{Model} & 
    \textbf{Accuracy} & 
    \textbf{F1$_m$} &  % 用下划线替代空格，更规范（可选）
    \textbf{Recall} & 
    \textbf{Precision} \\
    \hline
    BERT-PCL & 0.7972 & 0.7113 & 0.5294 & 0.5806 \\
    \hline
    GPT4 & 0.8252 & 0.7455 & 0.5588 & 0.6552 \\
    \hline
    VideoMAE & 0.7778 & 0.7090 & 0.5250 & 0.6176 \\
    \hline
    MultiPCL & 0.8309 & 0.7978 & 0.7632 & 0.6744 \\
    \hline
    CPCLDetector & 0.8382 & 0.8130 & 0.8878 & 0.7317 \\
    \hline
  \end{tabular}
  \label{table2}
  \vspace{-0.2cm}
\end{table}
\vspace{-0.2cm}
\subsection{Implementation Details and Evaluation Metrics}
The AdamW optimizer was used during training, with parameter settings: \(\beta_{1}=0.9\), \(\beta_{2}=0.999\), and a weight decay coefficient of 1e-3. The total number of training epochs was 150, and a cosine annealing learning rate scheduling strategy was adopted, with \(T_{max}\) set to 75 (half of the total number of epochs) and the minimum learning rate ( \(\eta_{min}\) ) set to 1e-6. To comprehensively evaluate the model performance (especially addressing the problem of PCL sample imbalance), four key metrics were adopted: accuracy, macro F1, recall, precision.
\vspace{-0.3cm}
\subsection{Experimental Results on the PCLMM}
Table 2 presents the results of different models on PCLMM. CPCLDetector outperforms the MultiPCL model(SOTA) across all metrics.
Accuracy increased by 0.73\% , macro F1 score rose by 1.52\% , recall improved significantly by 12.46\% , precision increased by 5.73\%. This indicates that the CPCLDetector model achieves higher overall prediction accuracy, stronger capability to capture implicit CPCL content, and fewer misjudgments. Paired t-tests yield p-values of 0.023 and 0.018 for Accuracy and Macro F1, both \textless 0.05.
\vspace{-0.3cm}
\begin{figure}[htbp]  % 位置参数：h(当前页)t(页顶)b(页底)p(单独页)，优先按 h→t→b→p 排序
  \centering  % 图片居中对齐
  % 插入图片：[width=比例\textwidth] 控制缩放，\includegraphics{文件名.格式}
  \includegraphics[width=0.5\textwidth]{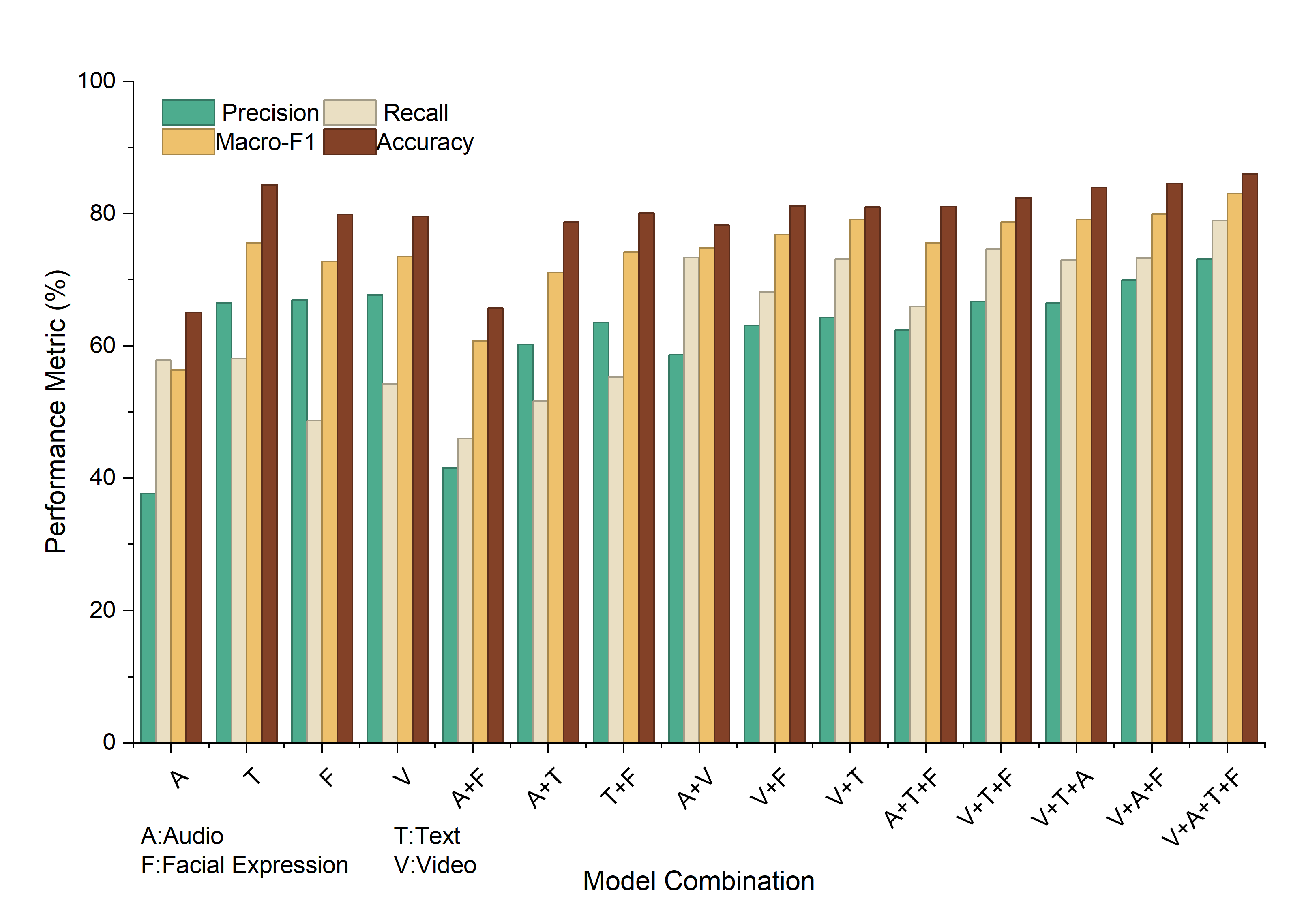}  
  % 图片标题（会自动编号，如 "Fig. 1"）
  \vspace{-0.8cm}  
  \caption{Performance comparison of different model combinations on PCLMM dataset.}  
  % 图片标签（用于文中引用，如 "\ref{fig:model_perf}"）
  \label{fig2}  
  \vspace{-0.3cm}
\end{figure}
\subsection{Experimental Results on the PCLMMPLUS Dataset}
As shown in Fig.2, CPCLDetector performs better on PCLMMPLUS.
Accuracy reaches 0.8603, an increase of 2.21\% compared to its performance on the PCLMM dataset;
Macro F1 score rises to 0.8305;
Recall maintains a high level of 0.8421;
Precision is 0.6667. Although this is lower than its performance on the PCLMM dataset, it remains within an acceptable range, and the model still effectively filters out most noise.
Benefiting from the increased diversity of PCL samples and the supplementation of comment information, the model maintains excellent balanced performance, confirming its robustness in detecting new types of PCL samples in the expanded dataset. Despite the fact that its precision on the PCLMMPLUS dataset is slightly lower than that on the PCLMM dataset, it is still acceptable, and the model can still effectively filter out most noise. One-way ANOVA yields p = 0.012 \textless 0.05 for Accuracy and Macro F1.  
\vspace{-0.4cm}
\begin{table}[htbp]
  \centering
  % 1. 缩小表格内字体（从默认 \normalsize 改为 \small，紧凑且不影响可读性）
  \small
  \caption{Ablation Study of CPCLDetector.}  % 表格标题（可按需修改）
  \label{table3}  % 标签用于文中引用（如 \ref{...}）
  % 2. 列格式优化：用 p{宽度} 替代 l，让 Model Variant 列自动换行，彻底解决宽度溢出
  %    p{6cm}：固定第一列宽度为 6cm（可根据你的页面宽度微调，如 5.5cm/6.5cm）
  %    |c|：保留中间两列居中对齐+竖线，符合原样式
  \begin{tabular}{p{4cm}|c|c}
    \hline
    % 3. 列标题换行：用 \makecell 拆分长标题，避免第一列过宽
    \textbf{\makecell{Model Variant}} & 
    \textbf{Accuracy} & 
    \textbf{\makecell{Accuracy Drop\\(vs. Full Model)}} \\
    \hline
    % 4. 内容对齐优化：第一列用 \centering 让文字居中（默认左对齐，居中更美观）
    \centering Full Model & 0.8603 & -- \\
    \hline
    \centering -- Comment Information Processing Module & 0.8451 & 1.52\% \\
    \hline
    % 5. 长文本自动换行：第一列宽度固定后，超长内容会自动折行，无需手动调整
    \centering -- Knowledge-Enhanced Sentiment Analysis Module & 0.8469 & 1.34\% \\
    \hline
    \centering Without Both Modules & 0.8361 & 2.42\% \\
    \hline
  \end{tabular}
\end{table}
\vspace{-0.8cm}
\subsection{Ablation Experiments}
To verify the value of the knowledge-enhanced comment content module, the research conducted ablation experiments on the PCLMMPLUS dataset (results are shown in Table 3).
After removing the comment information processing module, the accuracy decreased by 1.52\% ;
After removing knowledge-enhanced sentiment analysis module, the accuracy decreased by 1.34\% ;
The largest accuracy drop (2.42\%) occurred when both modules were removed simultaneously, confirming the synergistic contribution of the two modules. T-tests yield p-values of 0.015, 0.022, and 0.008 \textless 0.05 for removing comment, sentiment, and both modules, respectively.
\vspace{-0.6cm}
\section{Conclusion}
\label{sec:typestyle}

By constructing the PCLMMPLUS dataset and designing the CPCLDetector model, this study provides new insights for research on PCL detection in Chinese video scenarios.

The PCLMMPLUS dataset contains 831 samples, filling the gap of missing comment modalities in CPCL detection. The CPCLDetector model outperforms the baseline model on the PCLMM dataset with a particularly significant 12.46\% increase in recall.The research confirms that a knowledge-enhanced comment content module makes significant contributions. When the knowledge-enhanced comment content module is removed, the accuracy decreases by 2.42\%. Future research will further expand the scale of the dataset to more platforms and include richer levels of comments, while also exploring large-scale multi-modal models with alignment-free methods beyond text.

% References should be produced using the bibtex program from suitable
% BiBTeX files (here: strings, refs, manuals). The IEEEbib.bst bibliography
% style file from IEEE produces unsorted bibliography list.
% -------------------------------------------------------------------------
\bibliographystyle{IEEEbib}
\bibliography{main}

\begin{thebibliography}{10}

\bibitem{diao2023av}
Xingjian Diao, Ming Cheng, and Shitong Cheng,
\newblock ``Av-maskenhancer: Enhancing video representations through audio-visual masked autoencoder,''
\newblock in {\em 2023 IEEE 35th International Conference on Tools with Artificial Intelligence (ICTAI)}. IEEE, 2023, pp. 354--360.

\bibitem{li2023overview}
Bin Li, Yixuan Weng, Hu~Guo, Bin Sun, Shutao Li, Yuhao Luo, Mengyao Qi, Xufei Liu, Yuwei Han, Haiwen Liang, et~al.,
\newblock ``Overview of the nlpcc 2023 shared task: Chinese medical instructional video question answering,''
\newblock in {\em CCF International Conference on Natural Language Processing and Chinese Computing}. Springer, 2023, pp. 233--242.

\bibitem{li2024towards}
Shutao Li, Bin Li, Bin Sun, and Yixuan Weng,
\newblock ``Towards visual-prompt temporal answer grounding in instructional video,''
\newblock {\em IEEE transactions on pattern analysis and machine intelligence}, vol. 46, no. 12, pp. 8836--8853, 2024.

\bibitem{diao2025temporal}
Xingjian Diao, Chunhui Zhang, Weiyi Wu, Zhongyu Ouyang, Peijun Qing, Ming Cheng, Soroush Vosoughi, and Jiang Gui,
\newblock ``Temporal working memory: Query-guided segment refinement for enhanced multimodal understanding,''
\newblock {\em arXiv preprint arXiv:2502.06020}, 2025.

\bibitem{perez2020don}
Carla P{\'e}rez-Almendros, Luis Espinosa-Anke, and Steven Schockaert,
\newblock ``Don't patronize me! an annotated dataset with patronizing and condescending language towards vulnerable communities,''
\newblock {\em arXiv preprint arXiv:2011.08320}, 2020.

\bibitem{obadimu2019identifying}
Adewale Obadimu, Esther Mead, Muhammad~Nihal Hussain, and Nitin Agarwal,
\newblock ``Identifying toxicity within youtube video comment,''
\newblock in {\em International conference on social computing, Behavioral-cultural modeling and prediction and behavior representation in modeling and simulation}. Springer, 2019, pp. 214--223.

\bibitem{miyazaki2024impact}
Kunihiro Miyazaki, Takayuki Uchiba, Haewoon Kwak, Jisun An, and Kazutoshi Sasahara,
\newblock ``The impact of toxic trolling comments on anti-vaccine youtube videos,''
\newblock {\em Scientific Reports}, vol. 14, no. 1, pp. 5088, 2024.

\bibitem{Bertaglia_2024}
Thales Bertaglia, Catalina Goanta, and Adriana Iamnitchi,
\newblock ``The monetisation of toxicity: Analysing youtube content creators and controversy-driven engagement,''
\newblock in {\em 4th International Workshop on OPEN CHALLENGES IN ONLINE SOCIAL NETWORKS}. Sept. 2024, HT ’24, p. 1–9, ACM.

\bibitem{wang2025towards}
Hongbo Wang, Junyu Lu, Yan Han, Kai Ma, Liang Yang, and Hongfei Lin,
\newblock ``Towards patronizing and condescending language in chinese videos: A multimodal dataset and detector,''
\newblock in {\em ICASSP 2025-2025 IEEE International Conference on Acoustics, Speech and Signal Processing (ICASSP)}. IEEE, 2025, pp. 1--5.

\bibitem{wang2019talkdown}
Zijian Wang and Christopher Potts,
\newblock ``Talkdown: A corpus for condescension detection in context,''
\newblock {\em arXiv preprint arXiv:1909.11272}, 2019.

\bibitem{wang2023ccpc}
Hongbo Wang, Mingda Li, Junyu Lu, Liang Yang, Hebin Xia, and Hongfei Lin,
\newblock ``Ccpc: A hierarchical chinese corpus for patronizing and condescending language detection,''
\newblock in {\em CCF International Conference on Natural Language Processing and Chinese Computing}. Springer, 2023, pp. 640--652.

\bibitem{dosovitskiy2020image}
Alexey Dosovitskiy, Lucas Beyer, Alexander Kolesnikov, Dirk Weissenborn, Xiaohua Zhai, Thomas Unterthiner, Mostafa Dehghani, Matthias Minderer, Georg Heigold, Sylvain Gelly, et~al.,
\newblock ``An image is worth 16x16 words: Transformers for image recognition at scale,''
\newblock {\em arXiv preprint arXiv:2010.11929}, 2020.

\bibitem{zhang2016joint}
Kaipeng Zhang, Zhanpeng Zhang, Zhifeng Li, and Yu~Qiao,
\newblock ``Joint face detection and alignment using multitask cascaded convolutional networks,''
\newblock {\em IEEE signal processing letters}, vol. 23, no. 10, pp. 1499--1503, 2016.

\bibitem{huang2021facial}
Qionghao Huang, Changqin Huang, Xizhe Wang, and Fan Jiang,
\newblock ``Facial expression recognition with grid-wise attention and visual transformer,''
\newblock {\em Information Sciences}, vol. 580, pp. 35--54, 2021.

\bibitem{Ffmpeg}
P{\'e}FFmpeg Developers,
\newblock ``Ffmpeg.,''
\newblock {\em https://ffmpeg.org/}, 2024.

\bibitem{cui2020revisiting}
Yiming Cui, Wanxiang Che, Ting Liu, Bing Qin, Shijin Wang, and Guoping Hu,
\newblock ``Revisiting pre-trained models for chinese natural language processing,''
\newblock {\em arXiv preprint arXiv:2004.13922}, 2020.

\bibitem{li2025alignmamba}
Yan Li, Yifei Xing, Xiangyuan Lan, Xin Li, Haifeng Chen, and Dongmei Jiang,
\newblock ``Alignmamba: Enhancing multimodal mamba with local and global cross-modal alignment,''
\newblock in {\em Proceedings of the Computer Vision and Pattern Recognition Conference}, 2025, pp. 24774--24784.

\bibitem{ma2024less}
Fuyan Ma, Yiran He, Bin Sun, and Shutao Li,
\newblock ``Less is more: Adaptive feature selection and fusion for eye contact detection,''
\newblock in {\em Proceedings of the 32nd ACM International Conference on Multimedia}, 2024, pp. 11390--11396.

\bibitem{zhao2021knowledge}
Anping Zhao and Yu~Yu,
\newblock ``Knowledge-enabled bert for aspect-based sentiment analysis,''
\newblock {\em Knowledge-Based Systems}, vol. 227, pp. 107220, 2021.

\end{thebibliography}

\end{document}